\documentstyle[11pt,newpasp,twoside,epsfig,verbatim]{article}
\index{summary}
\index{instructions}
\index{template}

\def\edcomment#1{\iffalse\marginpar{\raggedright\sl#1\/}\else\relax\fi}
\reversemarginpar

\begin{document}
\title{
Simulations of pulsar wind formation
}
 \author{Anatoly Spitkovsky and Jonathan Arons}
\affil{Astronomy Dept., University of California, Berkeley, CA 94720, USA}

\begin{abstract}
We present initial results of the first self-consistent numerical model 
of the outer magnetosphere of a pulsar. By using the relativistic
``particle-in-cell'' method with special boundary conditions to represent 
plasma dynamics in 3D, we are able to follow magnetospheric plasma 
through the light cylinder into the wind zone for arbitrary magnetic 
inclination angles. For aligned rotators we confirm the ``disk-dome'' 
charge-separated structure of the magnetosphere and find that this 
configuration is unstable to a 3D nonaxisymmetric diocotron instability.
This instability allows plasma to move across the field lines and approach
the corotating Goldreich-Julian solution within several rotation periods. 
For oblique rotators formation of the spiral ``striped wind'' in the 
equatorial direction is demonstrated and the acceleration of the wind 
and its magnetization is discussed. We find that the wind properties vary
with stellar latitude; however, whether injection conditions at the pulsar
are responsible for the observed jet-equator geometry of Crab and Vela 
is currently under investigation. We also comment on the electrodynamics of 
the simulated magnetospheres, their current closure, and future simulations.
\end{abstract}

\section{Introduction}

Most of the pulsar wind models assume a relativistic MHD-type flow with various
degrees of symmetry (e.g., Kennel and Coroniti 1984,
Begelman and Li 1992), and are reasonably successful in explaining the overall
morphology and energetics of the surrounding plerion.  However, they run into
various difficulties under attempts to extrapolate the solution back towards
the pulsar. The most famous of these is the problem of the magnetization of the
wind ($\sigma$-paradox), which is the contradiction between the low ratio of
magnetic to kinetic energy ($\sigma \sim 10^{-3}$) at the wind-nebula interface
inferred from observations, {\it expected} strong magnetization near the pulsar
($\sigma \gg 1$), and the conservation of $\sigma$ in an ideal MHD flow.
Recent observations of the Crab and Vela with Chandra and HST have underscored
the importance of understanding the origins of the pulsar wind. Observations of
pulsar jets and orthogonal equatorial outflows suggest that rotation of the
neutron star is imprinted in the wind structure, possibly at a very early
stage. How and where acceleration and collimation of the wind occurs can be
reliably studied only by constructing a wind model that begins from the pulsar
itself.  The main difficulty in the way of theoretical progress on this front
is the lack of intuition about the phenomena occurring near a pulsar. For a
general case of an oblique rotator, and, as we show in this paper, even for an
aligned rotator, the problem is intrinsically three dimensional,
nonaxisymmetric, and time-dependent, or in other words ``too
complicated''. With observations unlikely to resolve light cylinder scales
($\sim 10^8$ cm) any time soon, our only resort is numerical simulations. In
this paper we describe our effort to simulate the formation of pulsar winds
from the first principles. In particular, we study the behavior of plasma that
is either emitted from or injected in the vicinity of a strongly magnetized
rotating conducting sphere with arbitrary inclination angles between the
magnetic and rotational axes. Despite its implied simplicity, this problem has
very interesting solutions that may at first appear strange and unusual. Hence,
in describing these preliminary results our emphasis is more on developing
intuition, rather than on direct relevance to pulsars and observational
implications. Such issues as well as more realistic models are currently being
investigated. The paper is organized as follows: in \S~2 we
describe the particle-in-cell method used in our simulations and the setup of
the problem, in \S~{3} and \S~{4} we describe numerical
experiments with aligned and oblique rotators, and in \S~{5} we
conclude and discuss further work.
 
\section{Numerical method} \label{secnumer}

We are interested in studying the outer magnetosphere of the pulsar --
the region around the light cylinder ($R_{LC}=c/\Omega$, where
$\Omega$ is the angular frequency of the star). The possible physical
conditions in this region put stringent demands on a successful
simulation method. Previous theoretical investigations suggest that
the method should be able to handle relativistic flows, vacuum gaps,
charge-separated configurations, space-charge limited flows and
countersteaming flows, as well as MHD flow when it is applicable. For
oblique rotators the light cylinder is a region of emission of
large-amplitude electromagnetic waves, so strong wave-plasma
interaction should be also addressable by the method. The requirement
of simulating counterstreaming, i.e. flows with multivalued velocity,
dispenses with the option of using fluid grid-based schemes. As our
method we therefore chose a particle-based ``Particle-In-Cell''
(PIC) algorithm. It is a fully relativistic multidimensional method
for self-consistent solution of Maxwell's equations in the presence of
plasma. 
The underlying
philosophy of PIC is the following: a plasma is represented
as a collection of macroparticles which carry charge and mass;
the macroparticles are moved by leap-frog integration of the relativistic
equations of motion with Lorentz force; the currents and charges
associated with the macroparticles are deposited on an Eulerian grid, on
which Maxwell's equations are also discretized; the fields are updated
(with particle currents and charges as sources) and extrapolated to
the position of the macroparticles to contribute to the Lorentz force for
the next time step.

The main advantage of using the PIC method for the pulsar wind problem is that
one does not have to impose restrictions on plasma behavior in advance. The
method can easily handle charge-separated flows and counterstreaming. It is
intrinsically a kinetic model, so particle acceleration can be studied.  The
ideal MHD condition $\vec{E}+(\vec{v}/c)\times \vec{B}=0$ can be achieved in
the appropriate regime, and has been observed in our tests. The disadvantage of
using such a fundamental plasma model is that multiple spatial and time scales
need to be resolved, which introduces constraints on the time step and spatial resolution 
(Birdsall and Langdon 1991).

We have modified the publically available 3D relativistic PIC code TRISTAN
(Buneman 1993) to simulate plasma dynamics in the magnetosphere of a magnetized
conducting rotating sphere.  As is well known, electric fields are induced on
the surface of such a body, and should serve as the boundary conditions for the
field solve.  In order to avoid the asymmetries introduced from the discretization
of a rotating sphere on a
Cartesian grid, we used the linearity of Maxwell's equations to represent the
field $\vec{E}_{total}$, which is used to advance the plasma particles, as a superposition:
$	\vec{E}_{total}=\vec{E}_{vacuum}+\vec{E}_{plasma}$,
where $\vec{E}_{vacuum}$ is the field of the ``vacuum rotator'', given by 
an analytic formula (Deutsch, 1955), and
$\vec{E}_{plasma}$ is the ``plasma'' field computed from the currents deposited
by macroparticles. The same decomposition is done for the magnetic field.  We
note that this is not a small amplitude expansion, this decomposition is exact,
and two components can be of equal magnitude.  A spherical surface at the
center of the domain acts a source and sink of macroparticles.  This surface
can represent the neutron star or a rigidly corotating inner magnetosphere
extending to a large fraction of the light cylinder (presently the radius of the sphere $a>0.1 R_{LC}$).
The outer walls of the computational domain in the simulation 
have radiation boundary condition for the fields and particles.

\begin{figure}[hbt]

\unitlength = 0.0011\textwidth
\hspace{10\unitlength}
\begin{picture}(200,200)(0,15)
\put(0,0){\makebox(200,200){ \epsfxsize=240\unitlength \epsfysize=200\unitlength
\epsffile{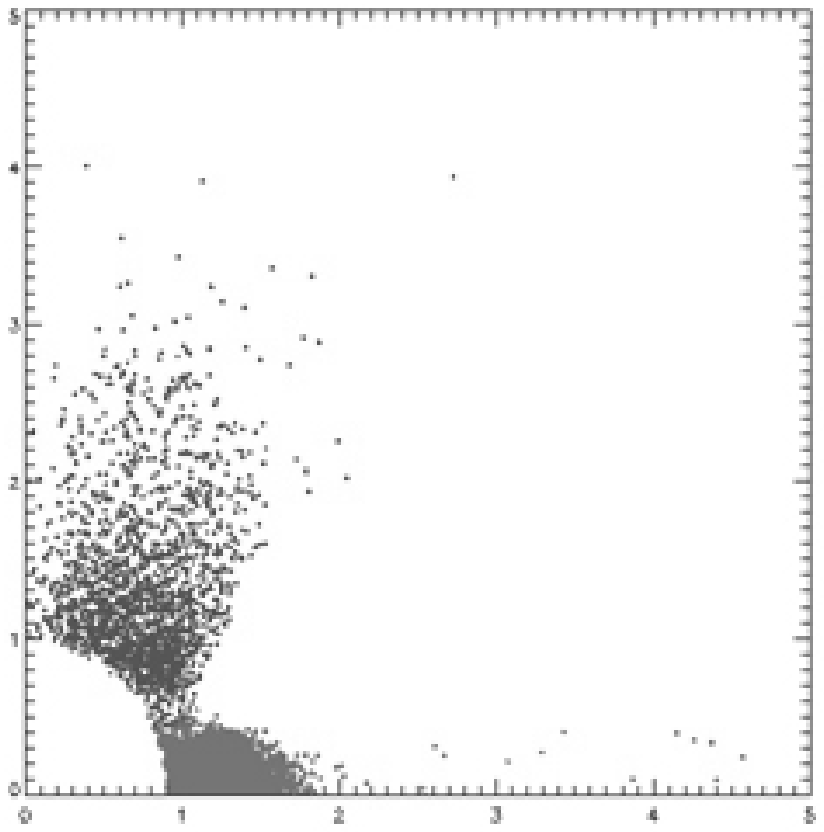}}}
\end{picture}
\hspace{3\unitlength}
\begin{picture}(200,200)(0,15)
\put(0,0){\makebox(200,200){\epsfxsize=200\unitlength \epsfysize=200\unitlength
\epsffile{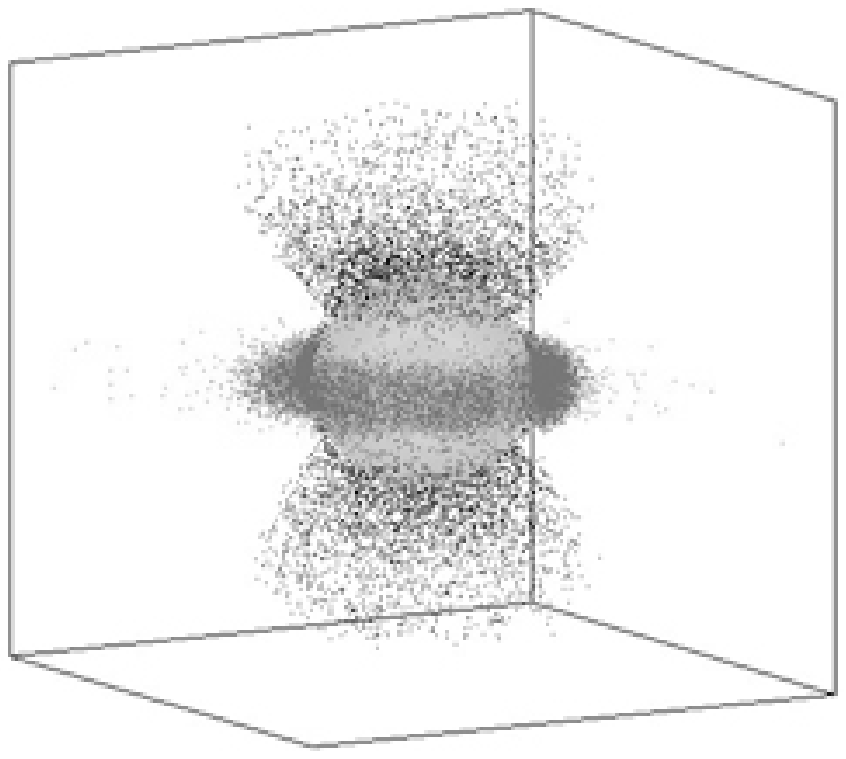}}}
\end{picture}
\hspace{1\unitlength}
\begin{picture}(200,200)(0,15)
\put(0,0){\makebox(200,200){\epsfxsize=200\unitlength \epsfysize=200\unitlength
\epsffile{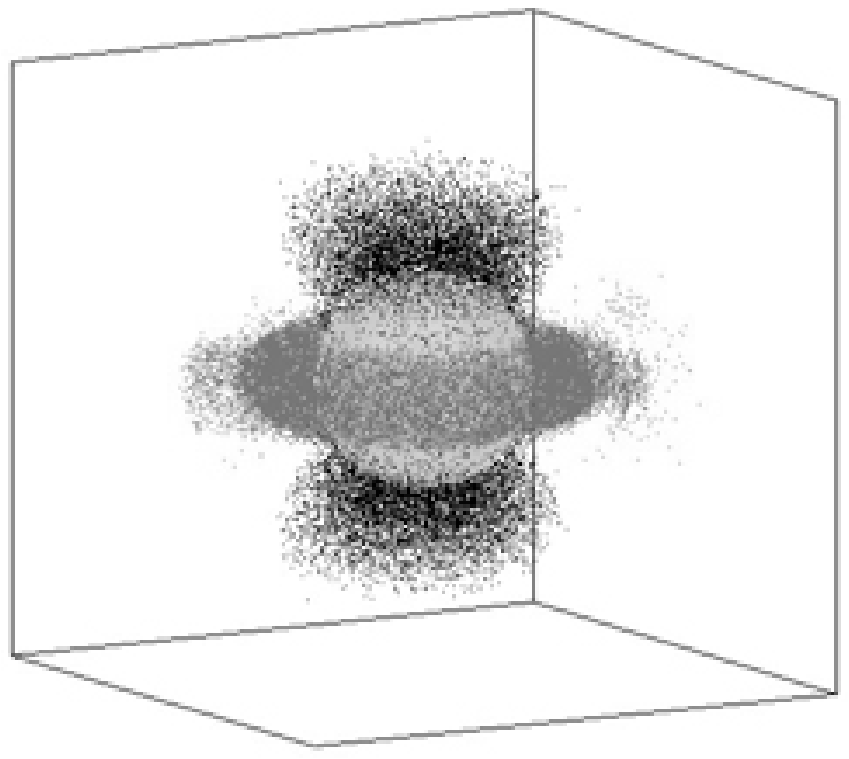}}}
\end{picture}
\hspace{1\unitlength}
\begin{picture}(200,200)(0,15)
\put(0,0){\makebox(200,200){\epsfxsize=200\unitlength \epsfysize=200\unitlength
\epsffile{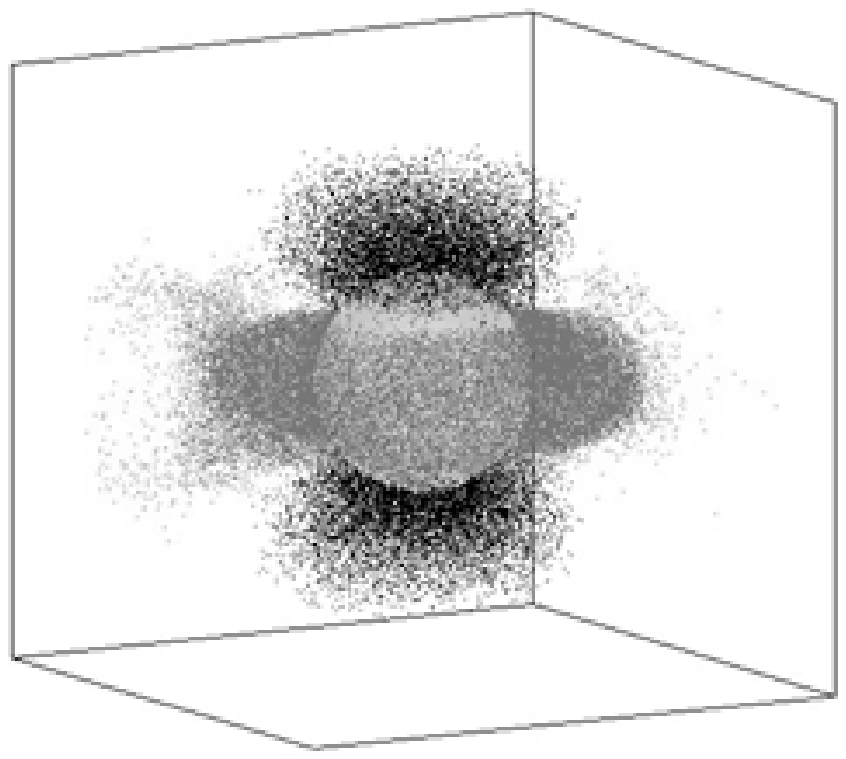}}}
\end{picture}


\vskip 0.1in
\hspace{10\unitlength}
\begin{picture}(200,200)(0,15)
\put(0,0){\makebox(200,200){ \epsfxsize=200\unitlength \epsfysize=200\unitlength
\epsffile{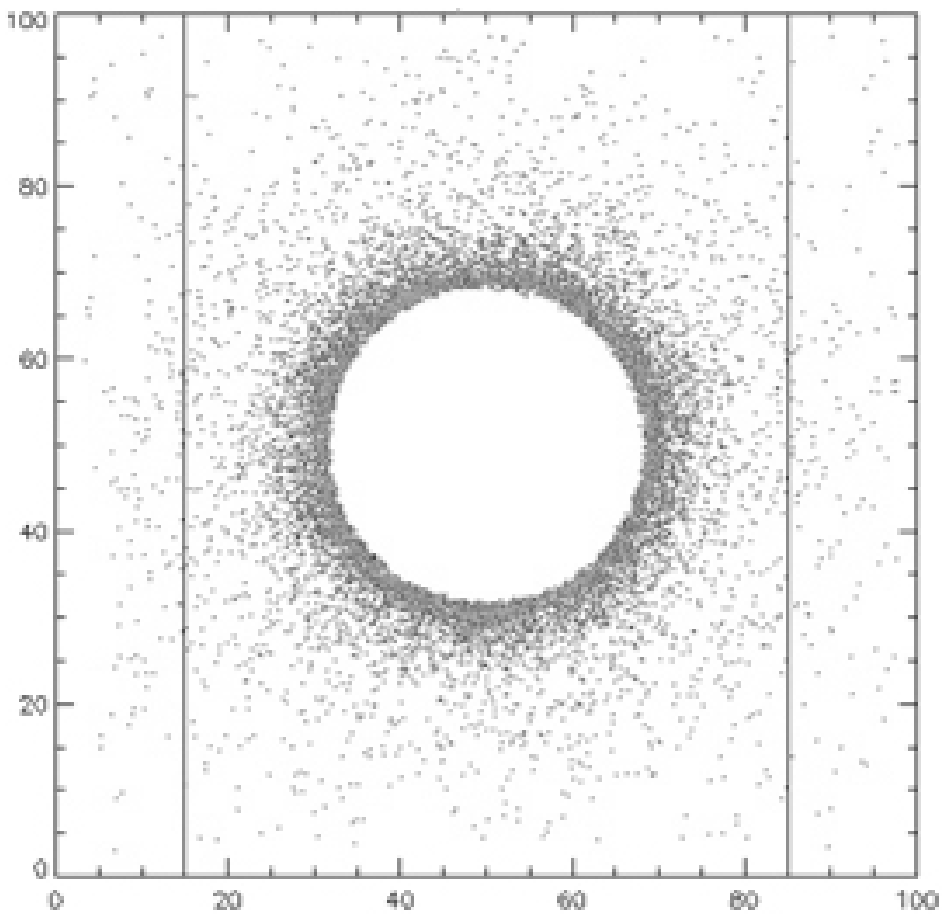}  
}}
\end{picture}
\hspace{3\unitlength}
\begin{picture}(200,200)(0,15)
\put(0,0){\makebox(200,200){\epsfxsize=200\unitlength \epsfysize=200\unitlength
\epsffile{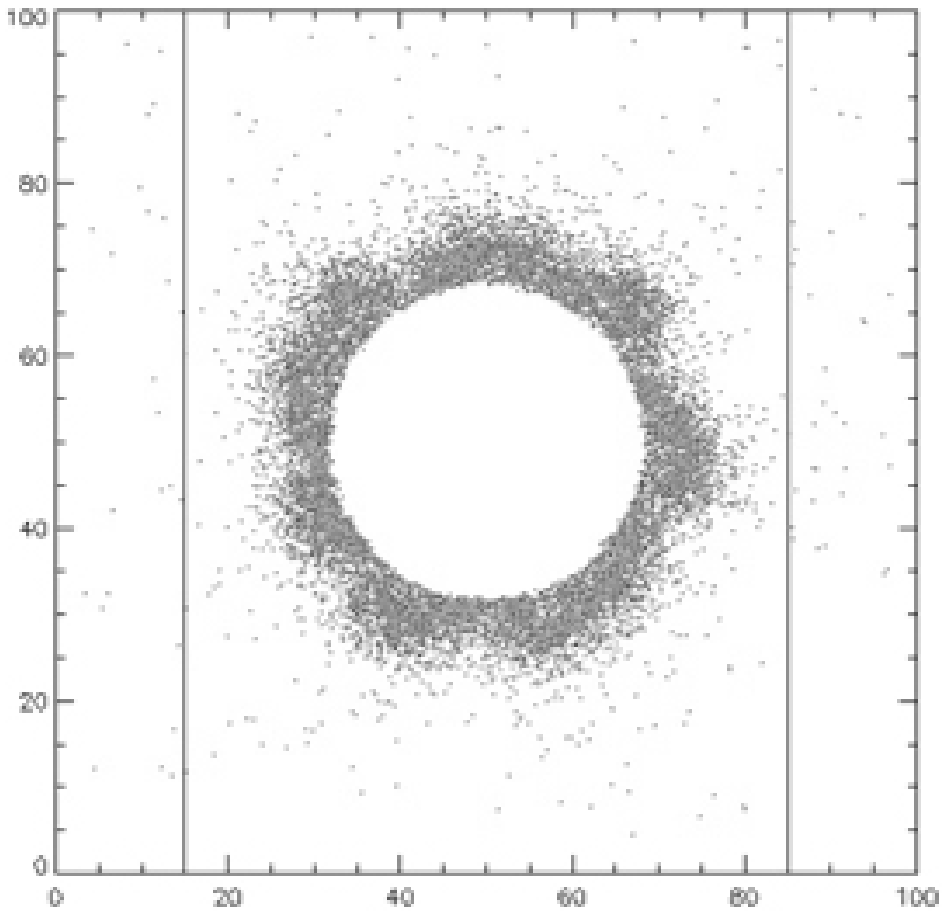} 
}}
\end{picture}
\hspace{1\unitlength}
\begin{picture}(200,200)(0,15)
\put(0,0){\makebox(200,200){\epsfxsize=200\unitlength \epsfysize=200\unitlength
\epsffile{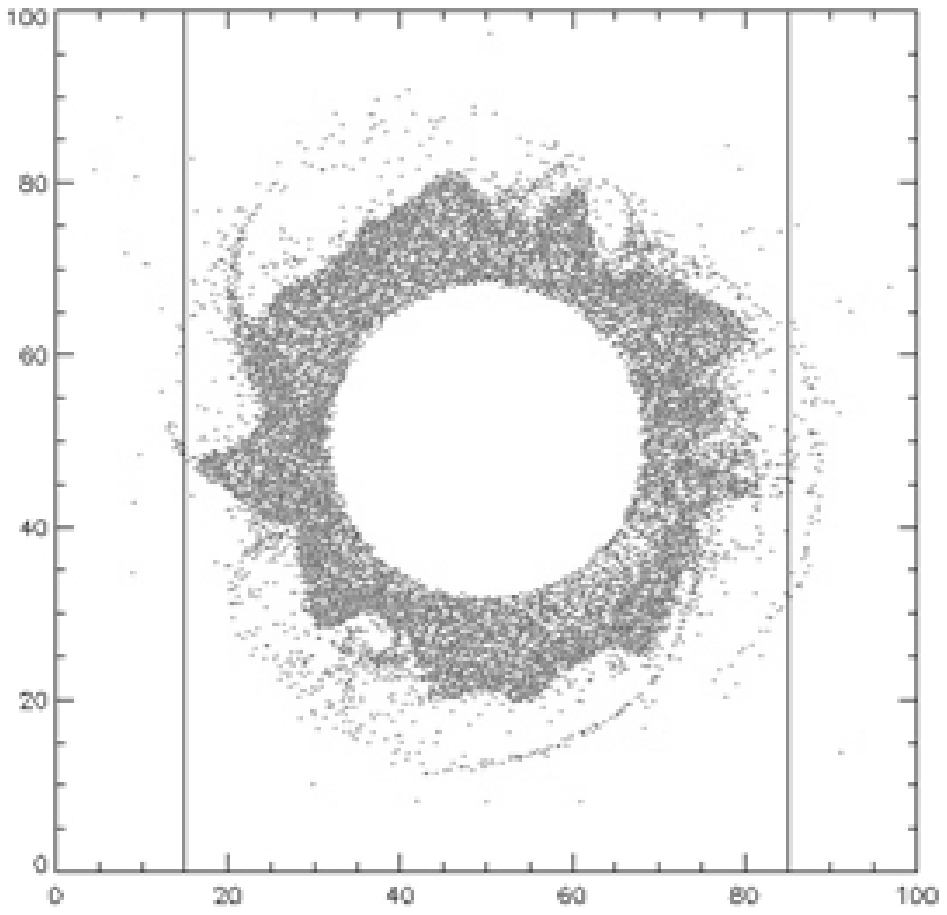}	
}}
\end{picture}
\hspace{1\unitlength}
\begin{picture}(200,200)(0,15)
\put(0,0){\makebox(200,200){\epsfxsize=200\unitlength \epsfysize=200\unitlength
\epsffile{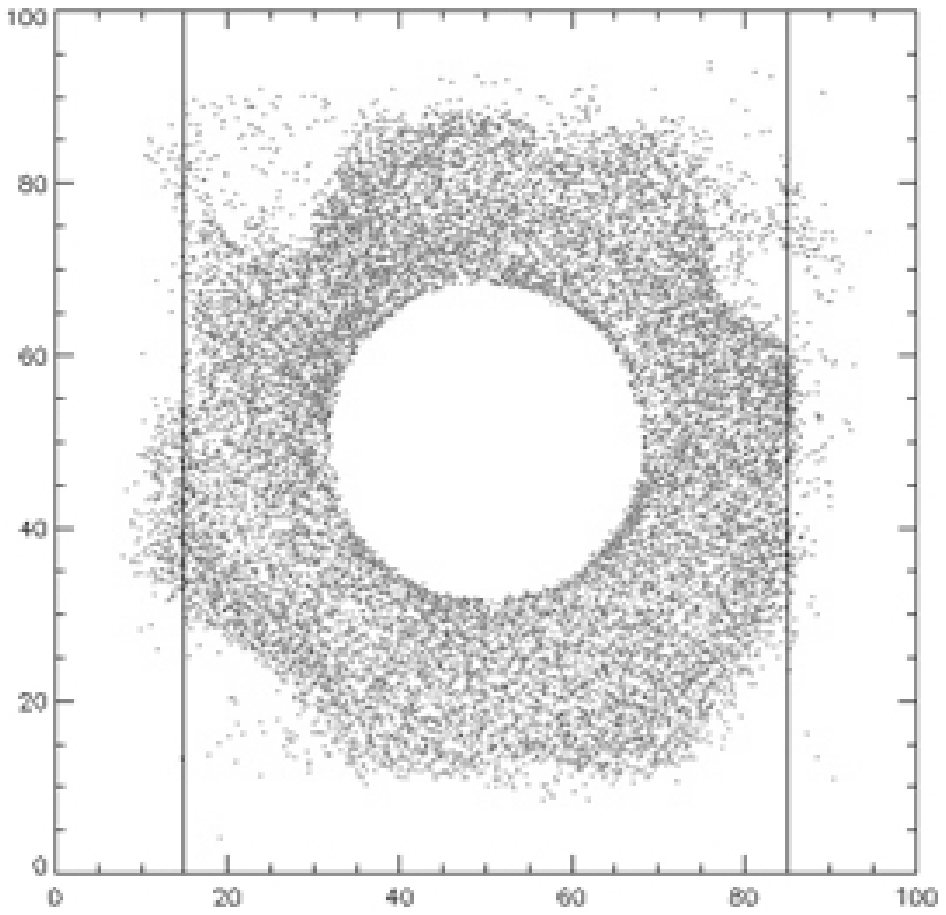} 
}}
\end{picture}
\hspace{1\unitlength}
\vskip 0.1in
\hspace{1\unitlength}
\begin{picture}(200,200)(0,15)
\put(0,0){\makebox(200,200){ \epsfxsize=200\unitlength \epsfysize=200\unitlength
\epsffile{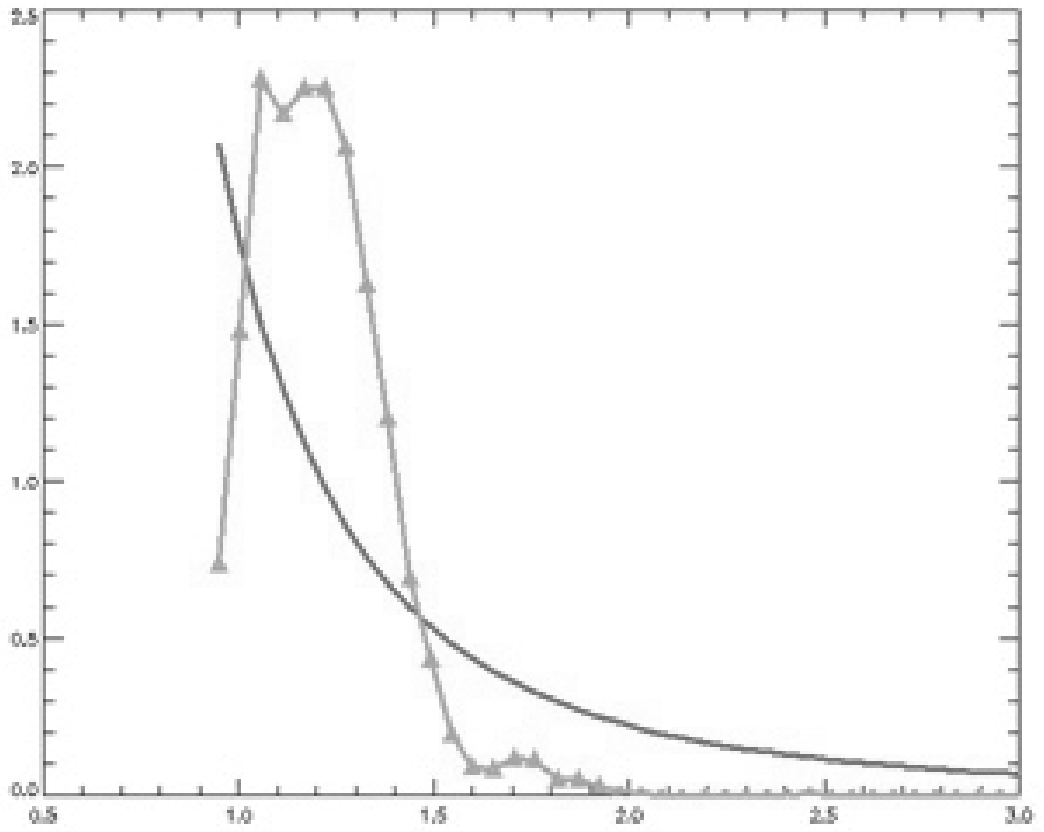}  
}}
\end{picture}
\hspace{3\unitlength}
\begin{picture}(200,200)(0,15)
\put(0,0){\makebox(200,200){\epsfxsize=200\unitlength \epsfysize=200\unitlength
\epsffile{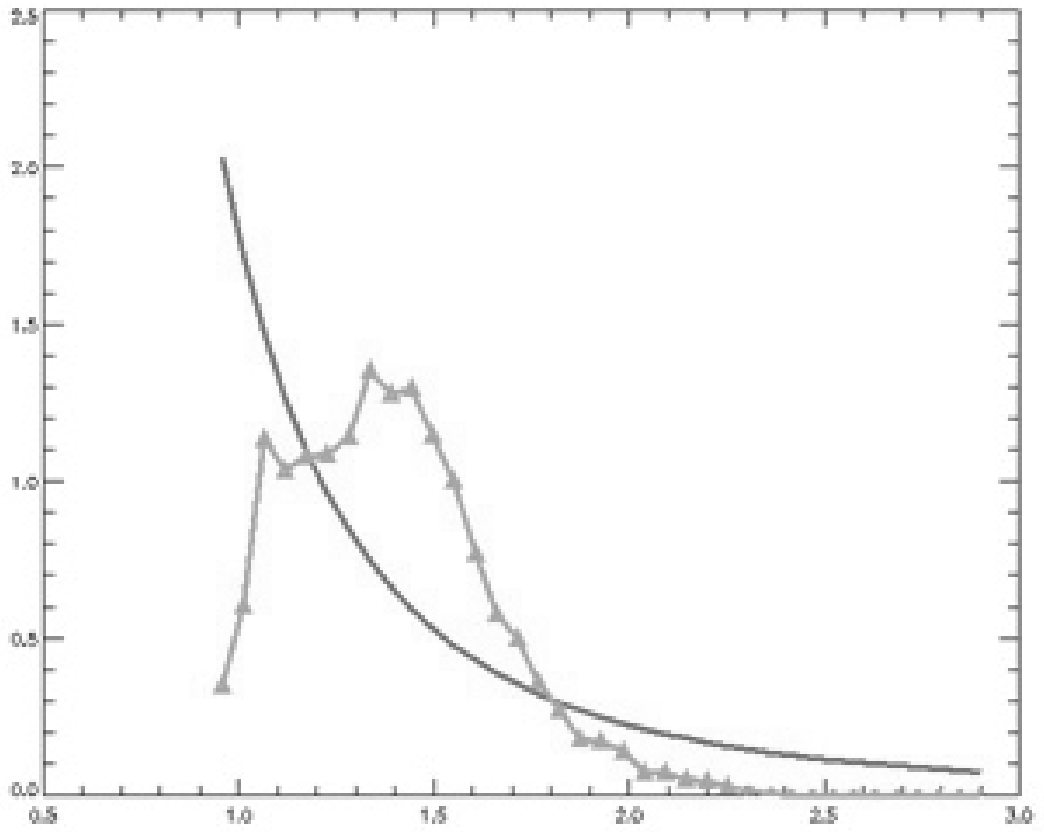} 
}}
\end{picture}
\hspace{1\unitlength}
\begin{picture}(200,200)(0,15)
\put(0,0){\makebox(200,200){\epsfxsize=200\unitlength \epsfysize=200\unitlength
\epsffile{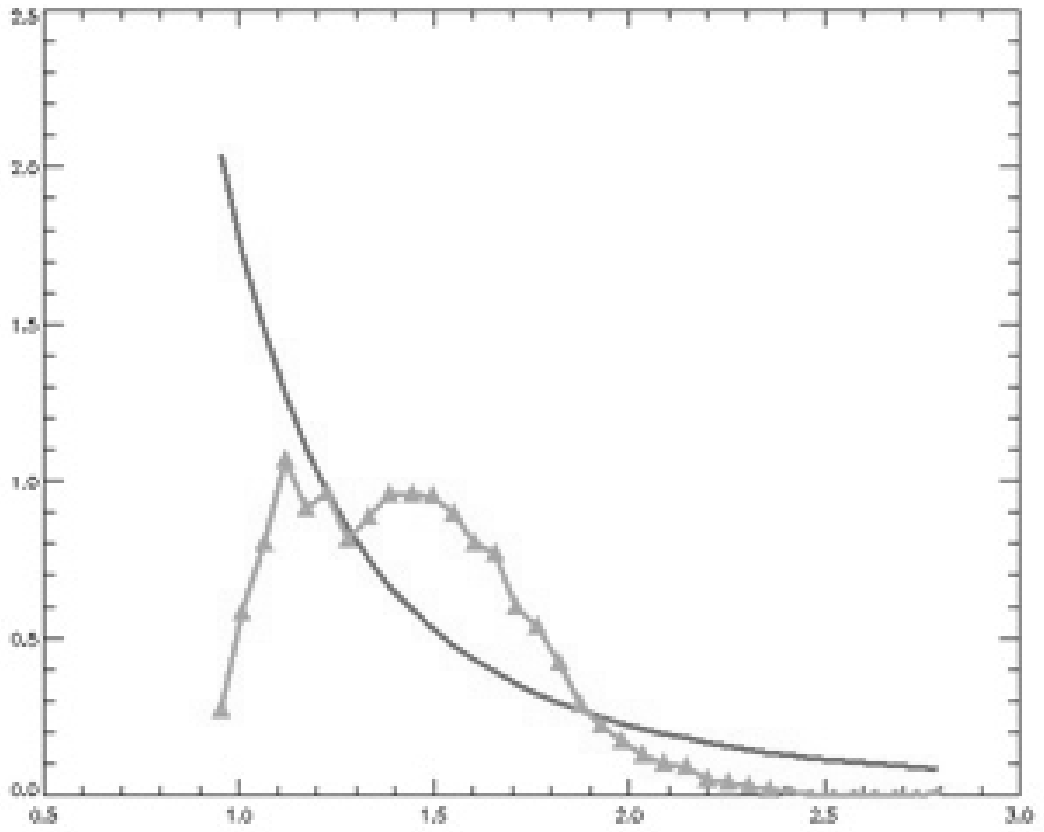}	
}}
\end{picture}
\hspace{1\unitlength}
\begin{picture}(200,200)(0,15)
\put(0,0){\makebox(200,200){\epsfxsize=200\unitlength \epsfysize=200\unitlength
\epsffile{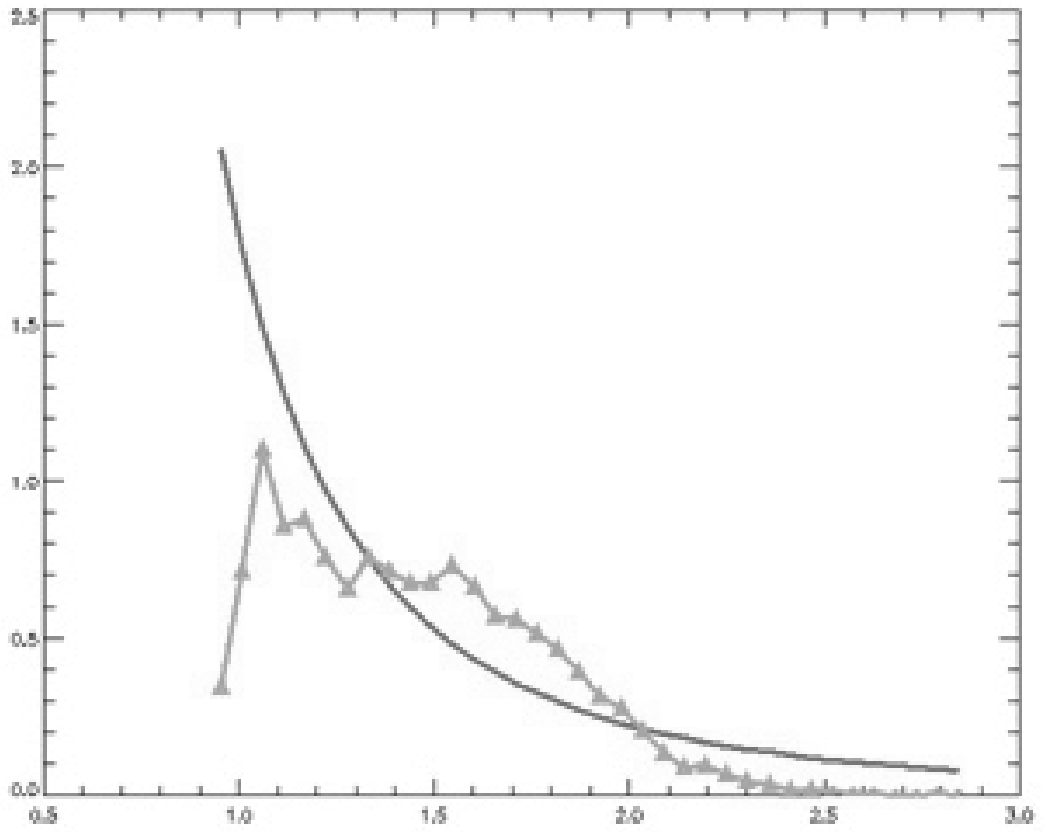} 
}}
\end{picture}
\hspace{1\unitlength}

\caption{\footnotesize {\it Top row:} Azimuthally averaged disk-dome configuration, and the 
3D view of the magnetosphere filling. {\it Middle row:} development of the diocotron instability
in the equatorial plane. Vertical lines represent the extent of the light cylinder for corotating plasma. 
{\it Bottom row:} Evolution of plasma density in the equatorial plane with the corresponding GJ solution (solid line). 
}
\end{figure}

\section{Aligned rotator}\label{secalign}
\vskip -.1 in
We begin by considering the case of magnetic axis being parallel to the
rotation axis.  There are no time-varying magnetic fields in this case, and the plasma
in the magnetosphere will know that the star is rotating only through the
electric fields of the unipolar inductor. Therefore, it is interesting to see
whether the magnetosphere can fill and reach corotation given a source of
plasma. It is natural to expect the plasma to be provided from the surface of
the star, as the induced electric field is likely to overwhelm any conceivable
work function. At the start of the simulation the rotating star has an induced
quadrupolar surface charge and a central monopole giving rise to the following
electric field components in spherical coordinates (e.g., Michel and Li 1999):
$E_r={2 \over 3} \phi_0 {a \over r^2}+\phi_0 {a^3 \over r^4} (1-3 \cos^2 \theta)$,
 $E_{\theta}=-\phi_0 {a^3 \over r^4} \sin 2 \theta$,
where $\phi_0 = \Omega B_0 a^2/c$, with $B_0$ magnetic field at the pole.
Plasma is introduced as electron-positron pairs at the stellar surface at each
time step.  The surface field extracts one sign of charge while driving the
other sign into the star where it is arrested.  This is equivalent to the
surface charge being torn off the star, eventually achieving a configuration
where the surface charge due to the vacuum fields is equal and opposite
to the surface charge induced by extracting the plasma particles, and the ${\bf
E}_{total}\cdot{\bf B}=0$ is satisfied on the surface (strictly speaking,
however, the charge left behind after extracting the plasma is not
distributed over the star; in this sense the star in the simulation is a
dielectric, rather than a conductor; see discussion in \S 5). The vacuum fields 
are such that electrons are extracted over the poles of the star
and positrons in the equatorial region. This creates a charge-separated plasma
with a vacuum region in between.  The quadrupolar field of this plasma
eventually counteracts the induced vacuum field at the surface, emission of new
charges is slowed down and the magnetosphere reaches a quasi-stable disk-dome
configuration as shown in Fig. 1. So far our results are in agreement with the previos
axisymmetric simulations using a different method (Krause-Polstorff and Michel
1984 (KPM1) and 1985 (KPM2); Smith et al 2001 (SMT)).  However, the 3D nature of
our simulation allows us to address new physics. 

The disk-dome configuration of Fig. 1 is not in rigid corotation with
the star. The equatorial disk consists of charges of only one sign and
has both velocity shear due to $E\times B$ drift in the total
vacuum+plasma field and a density gradient. These conditions are
favorable to the growth of the non-neutral plasma analog of the
Kelvin-Helmholtz instability -- the diocotron instability. This
instability proceeds by amplification of the azimuthal perturbations
in the charge density, and is illustrated in Fig. 1.  The
instability feeds on the energy of shear in the flow, and in the
process can move the plasma across the field lines (hence the
expanding radius of the disk in fig. 1). Indeed, an azimuthal
perturbation in charge density has a nonzero $E_{\theta}$ component of
the field, which results in a radial component of the $E\times B$ drift.

Such radial drifts represent the solution to the paradox of Holloway
(1973), who noted that there are certain regions in the
Goldreich-Julian (1969, GJ) solution that, if evacuated, would not be
directly replenished from the star with the correct sign of
charge. What probably happens in reality is that regions that are in
direct contact with the star will lose corotation and transfer charge
across the field lines due to the growing azimuthal perturbations in
the charge density. They would then be replenished directly
from the surface. Our simulations show similar behavior. The bottom panel of Fig. 1
displays the evolution of charge density in the equatorial plane as a
function of radius. As the instability develops, the density
tends to the Goldreich-Julian value. The agreement is not perfect,
however: the GJ solution implies an infinite extent of the charge. In
our case, the magnetosphere filled only up to the light
cylinder, with occasional filaments of charge being torn off after
moving past $R_{LC}$. In this sense, we did not get a self-consistent wind
with the aligned model. Part of the reason is the absence of the global
axial current that would significantly alter the magnetic topology.
We believe this can be modified with the more realistic boundary
conditions (see \S 5).  However, even now it is clear that the GJ
solution, corotation and the filling of the magnetosphere with plasma
are likely an end result of a robust dynamical process, which is
intrinsically three-dimensional in nature.

\section{Oblique rotator}\label{secobl}
With the three dimensional problem set up as described above, studying
oblique rotator amounts to just dialing the inclination angle in the
formula for the Deutsch field. For the test cases considered here we
also modified the plasma injection condition. At every step we injected
cold uniform pair plasma within a sphere of radius $R_{LC}$, in a
deliberately crude imitation of the pair formation process. If the
injected plasma density is larger than $\sim \rho_{GJ}$, the plasma has
a significant dynamical effect on the field, and after an initial
transient the system achieves a steady state (number of particles in
the domain is constant). The flow consists of two main regions: the
closed, quasi-corotating magnetosphere and the outflow (Fig. 2a). The first is
the region of the size of $R_{LC}$ encompassing the closed magnetic
field lines. Here the initially neutral plasma is separated with
excess positive charge around the equator and excess negative charge
around the poles, similar to the GJ solution. The
near-corotation is accomplished due to this charge separation and due to the
induced electric field that arises from time variation of the magnetic
field. Near the light cylinder the plasma is accelerated due to the
radiation pressure of the large amplitude electromagnetic wave and
forms an outflow.  An interesting feature of this wind is that it is
formed at the edge of the {\it closed} zone, and is mostly in the
sector around the rotational equator, where it has the characteristic spiral wave
pattern (Fig. 2b). Other regions of outflow are
also possible. In Fig. 2a we introduced an additional plasma source in
the polar flux tube for a $20^\circ$ rotator. The plasma acts to cancel the 
parallel electric fields and moves along the magnetic field tracing a cone around the 
rotation axis.


\vskip -.05in
\begin{figure}[hbt]
\unitlength = 0.0011\textwidth
\begin{center}
\hspace{1\unitlength}
a)
\hspace{25\unitlength}
\begin{picture}(200,200)(0,15)
\put(0,0){\makebox(200,200){ \epsfxsize=280\unitlength \epsfysize=280\unitlength
\epsffile{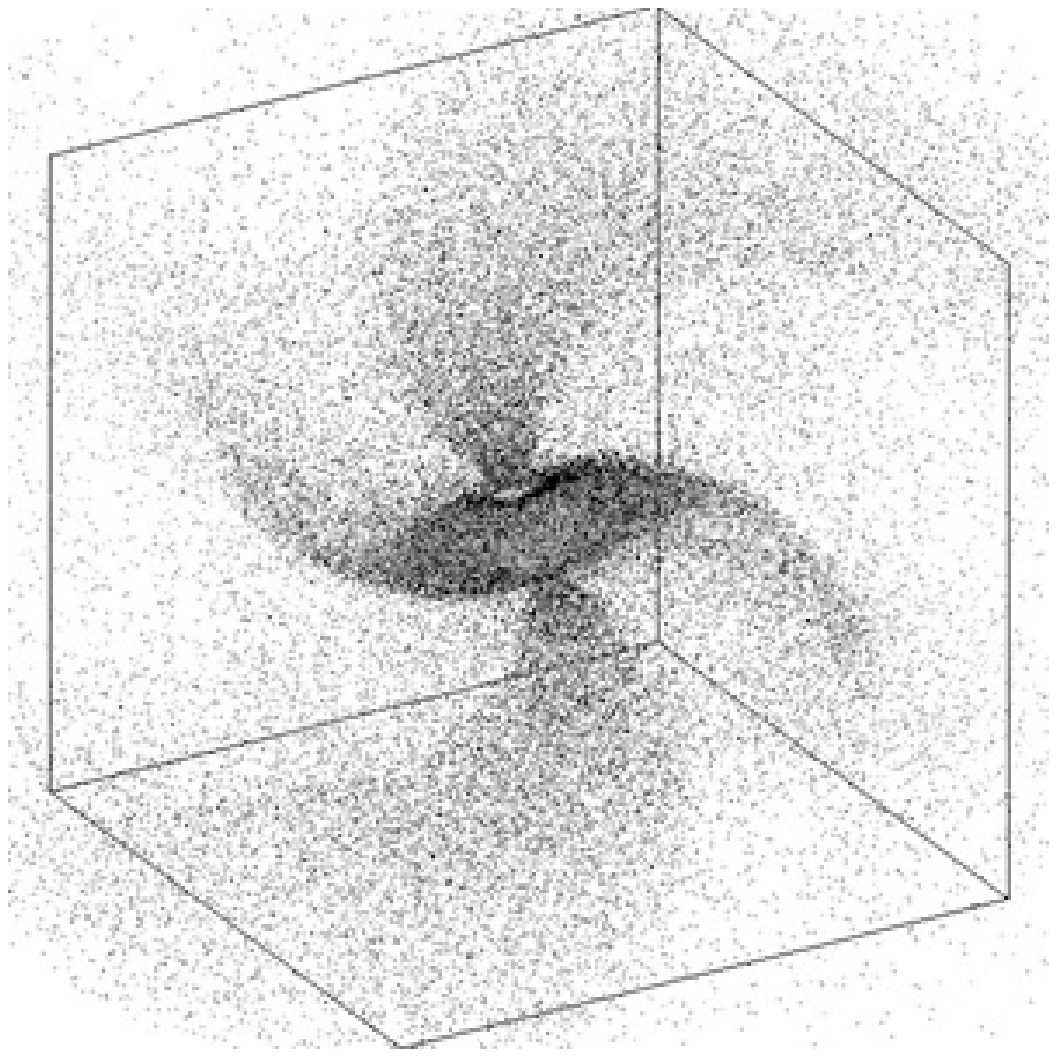}}}
\end{picture}
\hspace{60\unitlength}
b)
\hspace{25\unitlength}
\begin{picture}(200,200)(0,15)
\put(0,0){\makebox(200,200){\epsfxsize=280\unitlength \epsfysize=280\unitlength
\epsffile{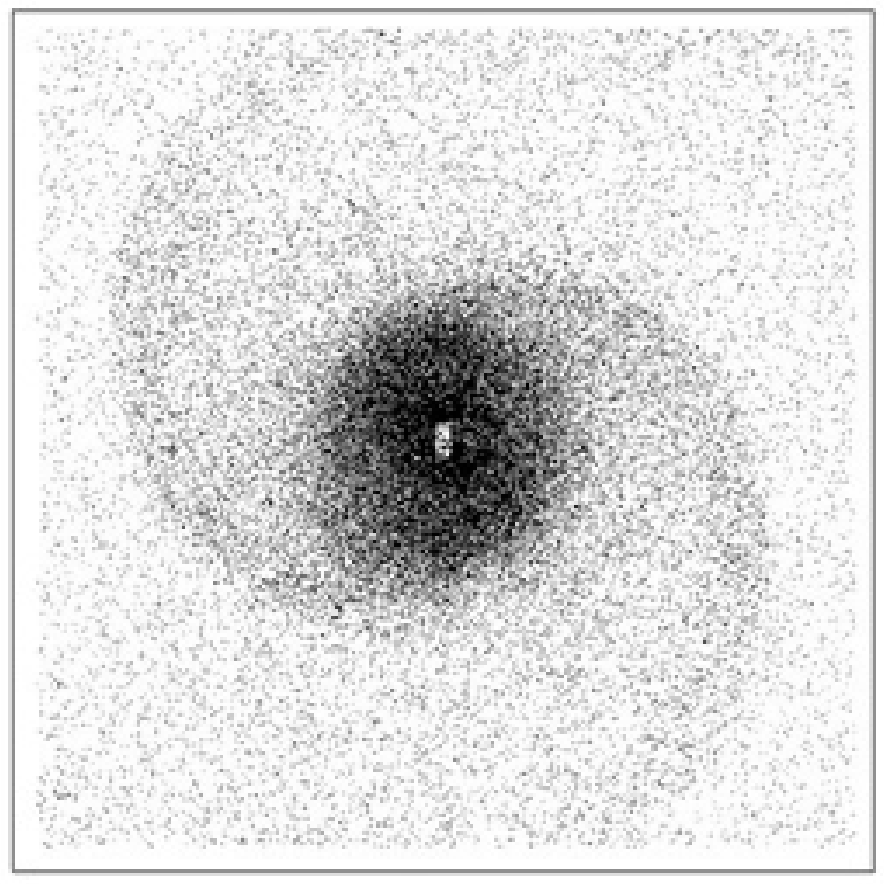}}}
\end{picture}
\hspace{1\unitlength}
\end{center}
\caption{ \footnotesize
a) Plasma distribution in a $20^\circ$ inclined rotator. Shown are the closed 
magnetosphere, equatorial outflow, and the polar flux tube; b) View of the equatorial plane 
illustrating the spiral wave formation. For the movie 
version of this figure please see 
http://astron.berkeley.edu/$\sim$anatoly/magnetosphere.html 
\label{fig2}}
\end{figure}

\section{Conclusion} \label{secconcl}
Our findings can be summarized as follows: pulsar magnetospheres {\it can} be filled
by surface emission with the help of a 3D diocotron instability; simple models
of oblique rotators {\it can} form relativistic winds; such winds have variation in 
properties with latitude, with equatorial flow being formed at the edge of the closed 
magnetosphere, and polar cap flow probably contributing in the direction of rotational axis. 
Much work remains to be done to quantify these qualitative results and
to understand the dependence of the wind geometry, acceleration and
collimation on the parameter regime. While tempting, it would be too
early to suggest that the equatorial and polar outflows observed in
these simulations are directly responsible for features in the Crab
and Vela. The models described in this paper are self-consistent in the sense
that the field and particle equations are solved simultaneously and
correctly. However, the models are not currently self-consistent with
respect to the boundary conditions: the induced quadrupole electric
fields are those from a rotating magnetized {\it conductor}, but as
far as the plasma is concerned the central body is a {\it dielectric}
with $\epsilon=1$. Both KPM1, KPM2 and SMT simulations suffer from the
same problem as well. This restriction does not allow currents to
close inside the star, and instead leads to charging of the central
body at the locations where the current is extracted and
returned. Consequently, any current flow in the system is
temporary. We are currently working on overcoming this
deficiency. 
The more realistic models with true central
conductors will be described in detail in an upcoming publication.


\begin{references}
\reference Begelman, M. C. and Li, Z.-Y. 1992, \apj, 397, 187
\reference Birdsall, C. K. and Langdon, A. B.  1991, ``Plasma Physics via Computer Simulation'' (Adam Hilger: Bristol)
\reference Buneman, O. 1993, in Computer Space Plasma Physics, ed. H. Matsumoto \& Y. Okura (Terra Scientific), 67
\reference Goldreich, P. and Julian, W. H. 1969, \apj, 160, 971 
\reference Deutsch, A. 1955, Annales d'Astrophysique, 18, 1
\reference Holloway, N. J., 1973, Nature Phys. Sci., 246, 6
\reference Kennel, C. F., \& Coroniti, F. V. 1984, \apj, 283, 694
\reference Krause-Polstorff, J. and Michel, F. C. 1984, \mnras, 213, 43 (KPM1)
\reference Krause-Polstorff, J. and Michel, F. C. 1985, A\&A, 144, 72 (KPM2)
\reference Michel, F. C. and Li, H. 1999, Phys. Reports,  318, 227
\reference Smith, I. A., Michel, F. C., Thacker, P. D. 2001, MNRAS, 322, 209 (SMT)



\end{references}
\end{document}